\documentclass[superscriptaddress,letterpaper,twocolumn,prb,showpacs]{revtex4-1}
\usepackage{amsmath, amsthm, amssymb}
\usepackage[pdftex]{graphicx}
\usepackage{dcolumn} 
\usepackage{bm} 

\usepackage{hyperref}
\hypersetup{backref,  
colorlinks=true,
allcolors=blue}

\makeatletter 
\gdef\@ptsize{2}
\let\@currsize\normalsize 
\makeatother 
\usepackage{setspace} 

\usepackage[usenames,dvipsnames]{color}

\begin{document}
\title{Observation of semi-localized dispersive states in the strongly correlated electron-doped ferromagnet Eu$_{1-x}$Gd$_{x}$O}

\author{D.E. Shai}
\affiliation{Laboratory of Atomic and Solid State Physics, Department of Physics, Cornell University, Ithaca, New York 14853, USA}
\author{M.H. Fischer}
\affiliation{Laboratory of Atomic and Solid State Physics, Department of Physics, Cornell University, Ithaca, New York 14853, USA}
\affiliation{Department of Condensed Matter Physics, Weizmann Institute of Science, Rehovot 7610001, Israel}
\author{A.J. Melville}
\affiliation{Department of Materials Science and Engineering, Cornell University, Ithaca, New York 14853, USA}
\author{E.J. Monkman}
\affiliation{Laboratory of Atomic and Solid State Physics, Department of Physics, Cornell University, Ithaca, New York 14853, USA}
\author{J.W. Harter}
\affiliation{Laboratory of Atomic and Solid State Physics, Department of Physics, Cornell University, Ithaca, New York 14853, USA}
\author{D.W. Shen}
\affiliation{Laboratory of Atomic and Solid State Physics, Department of Physics, Cornell University, Ithaca, New York 14853, USA}
\affiliation{State Key Laboratory of Functional Materials for Informatics, Shanghai Institute of Microsystem and
Information Technology (SIMIT), Chinese Academy of Sciences, Shanghai 200050, China}
\author{D.G. Schlom}  
\affiliation{Department of Materials Science and Engineering, Cornell University, Ithaca, New York 14853, USA}
\affiliation{Kavli Institute at Cornell for Nanoscale Science, Ithaca, New York 14853, USA}
\author{M.J. Lawler}
\affiliation{Department of Physics, Binghamton University, Binghamton, New York 13902, USA}
\affiliation{Laboratory of Atomic and Solid State Physics, Department of Physics, Cornell University, Ithaca, New York 14853, USA}
\author{E.-A. Kim}
\affiliation{Laboratory of Atomic and Solid State Physics, Department of Physics, Cornell University, Ithaca, New York 14853, USA}
\author{K.M. Shen}
\email[Author to whom correspondence should be addressed: ]{kmshen@cornell.edu}
\affiliation{Laboratory of Atomic and Solid State Physics, Department of Physics, Cornell University, Ithaca, New York 14853, USA}
\affiliation{Kavli Institute at Cornell for Nanoscale Science, Ithaca, New York 14853, USA}

\begin{abstract}
Chemical substitution plays a key role in controlling the electronic and magnetic properties of complex materials. For instance, in EuO, carrier doping can induce a spin-polarized metallic state, colossal magnetoresistance, and significantly enhance the Curie temperature. Here, we employ a combination of molecular-beam epitaxy, angle-resolved photoemission spectroscopy, and an effective model calculation to investigate and understand how semi-localized states evolve in lightly electron doped Eu$_{1-x}$Gd$_{x}$O above the ferromagnetic Curie temperature. Our studies reveal a characteristic length scale for the spatial extent of the donor wavefunctions which remains constant as a function of doping, consistent with recent tunneling studies of doped EuO. Our work sheds light on the nature of the semiconductor-to-metal transition in Eu$_{1-x}$Gd$_{x}$O and should be generally applicable for doped complex oxides. 
\end{abstract}


\maketitle


Chemical substitution is one of the most common pathways for controlling the properties of electronic materials, from modifying the resistivity of semiconductors,\cite{Shklovskii:1984} realizing high-temperature superconductivity,\cite{kastner:1998} or accessing nanoscale spin, charge, or orbitally ordered states.~\cite{imada:1998} When stoichiometric, the binary oxide EuO is a dense ferromagnet ($S$ = 7/2) with a half-filled, $4f^{7}$ shell. Carrier doping induces a number of remarkable properties, including a metal-insulator transition $(\Delta \rho/ \rho_{0} \approx 10^{13})$,\cite{petrich:1971, penney:1972} colossal magnetoresistance $(\Delta \rho/ \rho_{0} \approx 10^{6})$,\cite{shapira:1973} an enhancement of the Curie temperature ($T_{\rm C}$),\cite{Shafer:1968} and highly spin-polarized carriers ($>$ 90\%),\cite{steeneken:2002, schmehl:2007} leading many to explore this material for spintronic applications e.g., a spin valve.~\cite{santos:2004} Carrier doping can be achieved in EuO by a number of methods, including oxygen vacancies or the substitution of Eu with Gd, leading to the introduction of $n$-type donors which ultimately results in a degenerately doped phase at high doping. Nonetheless, how the electronic structure of this system evolves in the low doping regime remains an open question. 

In a simple band-insulator scenario, one might expect metallic behavior from doping even an infinitesimal amount of carriers, although in reality, an insulating ground state persists for some finite range of doping. Metallic behavior is not achieved until enough donors are introduced such that their wave functions begin to overlap to the point where they form a degenerate state. In typical Group IV doped semiconductors such as Si:P, the Bohr radius of the donors can be many tens of angstroms, meaning that this semiconductor-to-metal transition occurs at very low carrier concentrations of $n \approx 10^{18}$ cm$^{-3}$. The low concentration makes it difficult to probe this regime using direct probes of the electronic structure such as angle-resolved photoemission spectroscopy (ARPES), since the photoemission intensity is proportional to the total number of states. Furthermore, it remains unclear whether this phenomenology from conventional semiconductors can also be applied to strongly correlated magnetic materials such as Eu$_{1-x}$Gd$_{x}$O. In many oxides including EuO, this transition occurs at significantly higher densities, typically on the order of $n \approx 10^{20}$ cm$^{-3}$, making this regime accessible by ARPES and related techniques. Despite the strong Coulomb interactions, which split the $4f$ states, above $T_{\rm C}$ one can approximate EuO as a semiconductor with a band gap ($\sim1.1$ eV) separating an occupied Eu $4f$ valence band from an unoccupied Eu $5d$ conduction band. 

\begin{figure*}[!t]
	\centering
	\includegraphics[width=\linewidth]{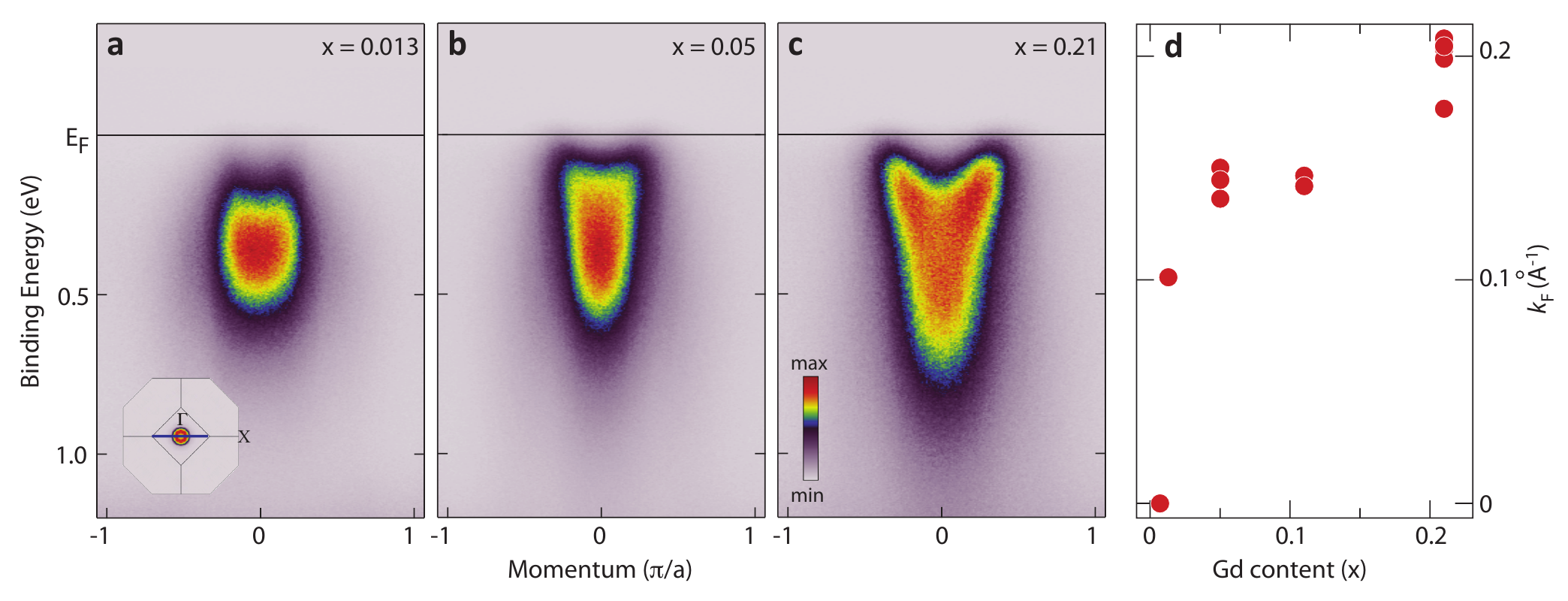}
	\caption{\label{fig:SpectralFunctionEvolution} (a)-(c) ARPES dispersions for Eu$_{1-x}$Gd$_{x}$O taken at constant temperature $T=140$ K around $\Gamma$ while varying $x$. The inset in (a) is the Fermi surface for Eu$_{0.95}$Gd$_{0.05}$O, reproduced from Ref. \onlinecite{shai:2012}, illustrating where in $k-$space the cut is taken. (d) Evolution of the Fermi wavevector $k_\mathrm{F}$ with $x$.}
\end{figure*}


Here, we report the evolution of the electronic structure at low carrier dopings in electron-doped Eu$_{1-x}$Gd$_{x}$O, using a combination of reactive oxide molecular beam epitaxy, ARPES, and model Hamiltonian calculations to explain our observations. Thin films of (001) Eu$_{1-x}$Gd$_{x}$O were grown on (110) YAlO$_{3}$ single crystal substrates by molecular beam epitaxy in both a dual chamber Veeco 930 and GEN10 system. The substrate was heated to 400 $^{\circ}$C during growth, resulting in adsorption-controlled growth of EuO, which minimizes oxygen nonstoichiometry.~\cite{ulbricht:2008,sutarto:2009} The Eu flux was $1.1\times10^{14}$ atoms/cm$^{2}$s, and the Gd flux was varied to achieve different doping levels ($x$). Film quality was monitored during growth using reflection high-energy electron diffraction. Following growth, the films were immediately transferred to the ARPES system without breaking ultra-high vacuum ($2\times10^{-10}$ torr). ARPES measurements were performed using a VG Scienta R4000 spectrometer and He I$\alpha$ photons ($h\nu$ = 21.2 eV) The instrumental energy resolution was $\Delta E=25$ meV and the base pressure typically was better than $6\times10^{-11}$ torr. 


We begin our discussion by analyzing the doping-dependent ARPES spectra for Eu$_{1-x}$Gd$_{x}$O ($x$ = 0.013, 0.05, and 0.21) at a fixed temperature, $T=140$ K, above $T_{\rm C}$ for all samples. Prior work\cite{shai:2012} has revealed that above $T_{\rm C}$, carriers introduced by rare earth doping result in the formation of a circular electron pocket near the Brillouin Zone center $\Gamma$, ($k_{x}$, $k_{y}$) = 0, shown in Fig. \ref{fig:SpectralFunctionEvolution}(a, inset). At the lowest doping levels ($x$ = 0.013, which sits on the boundary of the insulator-to-metal transition), the spectral weight near $E_{\rm F}$ is comprised of a nearly featureless patch of intensity; samples with doping levels below $x$ = 0.013 showed no observable weight near $E_{\rm F}$ and charged up electrostatically above $T_{\rm C}$ during measurements, suggesting these states are bulk-derived and do not arise from surface states or surface accumulation layers. Upon moving to higher doping levels, the spectral features evolve to a dispersive band which can be clearly observed for $x$ = 0.21, well in the metallic regime. The evolution of $k_{\mathrm{F}}$ with $x$ [Fig. \ref{fig:SpectralFunctionEvolution}(d)] shows that the size of this pocket increases with Gd content, consistent also with higher carrier densities observed by Hall effect measurements at 4.2 K.~\cite{mairoser:2010}


While our ARPES measurements characterize how these doping-induced states evolve with $x$, the nature and origin of these states is unclear. As noted in Ref.~\onlinecite{shai:2012}, a rigid band shift alone cannot explain the appearance of a pocket around the $\Gamma$ point, as the conduction band minimum should be at the Brillouin zone boundary ($X$ point). Major advances have been made in calculating the effects of disorder on the band structure in the iron-based superconductors and other systems by employing a combination of supercell band structure calculations coupled with an unfolding procedure.~\cite{ku:2010, berlijn:2012} Due to the correlated nature of EuO and the relatively low doping limit that we are investigating, we instead employ a qualitative model Hamiltonian considering dilute impurity states~\cite{herng:2010} that weakly overlap. In the following, we describe the impurity states by means of a tight-binding model, 
\begin{equation}
    \mathcal{H} = \sum_{i\neq j}t_{ij} d_{i}^\dag d^{\phantom{\dag}}_{j} + \sum_i \epsilon_i d^\dag_{i} d^{\phantom{\dag}}_{i} \equiv \sum_{i, j}\hat{\mathcal{H}}_{ij} d_{i}^\dag d^{\phantom{\dag}}_{j},
    \label{eq:tb}
\end{equation}
where the sums run over the $n=xN$ (random) impurity sites with random on-site potential $\epsilon_i$ and $t_{ij}$ are the hopping integrals. Note that the operator $d^\dag_{i}$ creates an electron at the impurity site $i$ in a spatially-extended impurity-state wavefunction $\phi_i(\vec{r}-\vec{r}_i)$. Using the Fourier transform $\tilde{\phi}_i(\vec{k})$ of these wave functions, we can thus express the electron Green's function in terms of the tight-binding propagator $g(i,j,\omega) = [(\omega - \hat{\mathcal{H}})^{-1}]_{ij}$, 
\begin{equation}
    G^{(\rm ret)}(\vec{k}, \vec{k}, \omega) = \sum_{i,j}\tilde{\phi}_i(\vec{k})\tilde{\phi}_j(\vec{k})e^{i\vec{k}\cdot(\vec{r}_i-\vec{r}_j)}g(i,j,\omega^+).
\end{equation}
Finally, the single-electron spectral density measured by ARPES can be calculated as
\begin{equation}
    \mathcal{A}(\vec{k},\omega) = -\frac{1}{\pi}\mathrm{Im}\ G^{(\rm ret)}(\vec{k},\vec{k},\omega).
\end{equation} 

For the case of localized impurities, i.e., $t_{ij} \equiv 0$, the spectral function is the sum of the individual impurity spectral functions, $S_i(\vec{k}, \omega) = |\tilde{\phi}_i(\vec{k})|^2 \delta(\omega - \varepsilon_i)$, such that the width of the ARPES signal in momentum space is given by the inverse of the size of the impurity state in real space. For the semi-localized case, where the impurity wave functions overlap, we can approximate the spectral function by

\begin{equation}
  S(\vec{k}, \omega)\approx |\tilde{\phi}(\vec{k})|^2\Big[-\frac{1}{\pi}\mathrm{Im}\Big(\sum_{i,j}e^{i\vec{k}\cdot(\vec{r}_i-\vec{r}_j)}g(i,j,\omega^+)\Big)\Big]
    \label{eq:stot}
\end{equation}
assuming that all of the impurity wave functions are approximately the same. The electron Green's function is thus given by the Fourier transform of the tight-binding propagator, with $|\tilde{\phi}(\vec{k})|^2$ as an envelope function. 

\begin{figure}
    \includegraphics[width=0.5\textwidth]{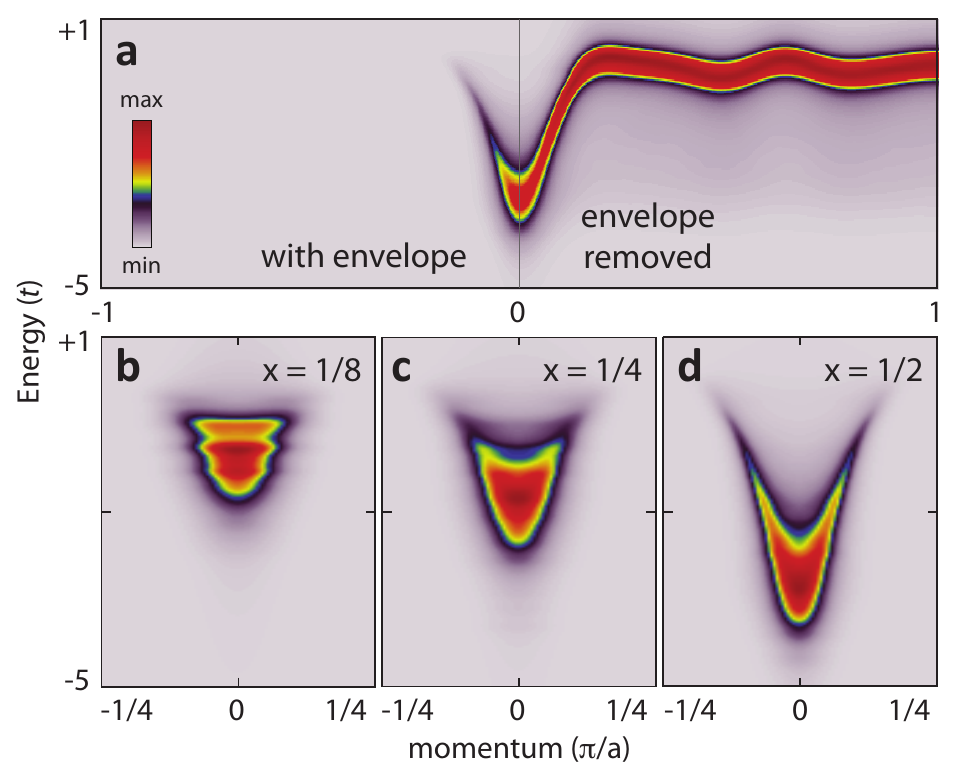}
    \caption{\label{fig:simulation} (a) Simulated spectral function for a one-dimensional system ($N=1000$ sites) consisting only of spatially extended impurity states with the Gaussian envelope removed to highlight the impurity band dispersion. (b-d) Isolated impurity band spectral functions calculated for doping values of 1/8, 1/4, and 1/2.  The range of the color scales for each panel in has been individually normalized to make the impurity bands more visible.}
\end{figure}

Figure~\ref{fig:simulation} exemplifies the resulting effect on the single-electron spectral density by using a one-dimensional model. Note that for simplicity, we use Gaussian impurity wave functions with a width $r_0$. A key result from this model calculation is that the envelope of the spectral intensity observed in ARPES for the doped carriers is set by the Fourier transform of this Gaussian envelope and is doping independent. In addition, we allow for an extended hopping of the form $t_{ij} = t \exp[-(\vec{r}_i-\vec{r}_j)^2/2r_0^2]$, with the hopping energy $t$ setting the energy scale. This hopping accounts for the extended nature of the impurity states and leads to an appreciable dispersion around the $\Gamma$ point, even at low doping, consistent with the location in momentum space of the semi-localized states observed in experiment in Fig. \ref{fig:SpectralFunctionEvolution}. The spectral functions are averaged over 100 random impurity arrangements, random on-site energies $\epsilon_i/t\in [-0.5-0.05, -0.5+0.05]$ and $r_0=4a$ shown both with and without this envelope, and with the envelope at three different doping values.  Our calculation illustrates the formation of a dispersive impurity band, together with the momentum-dependent suppression of intensity when moving away from $\Gamma$, which is parameterized by the spatial extent of the impurity state, $r_0$. 


\begin{figure*}
	\centering
	\includegraphics[width=1\textwidth]{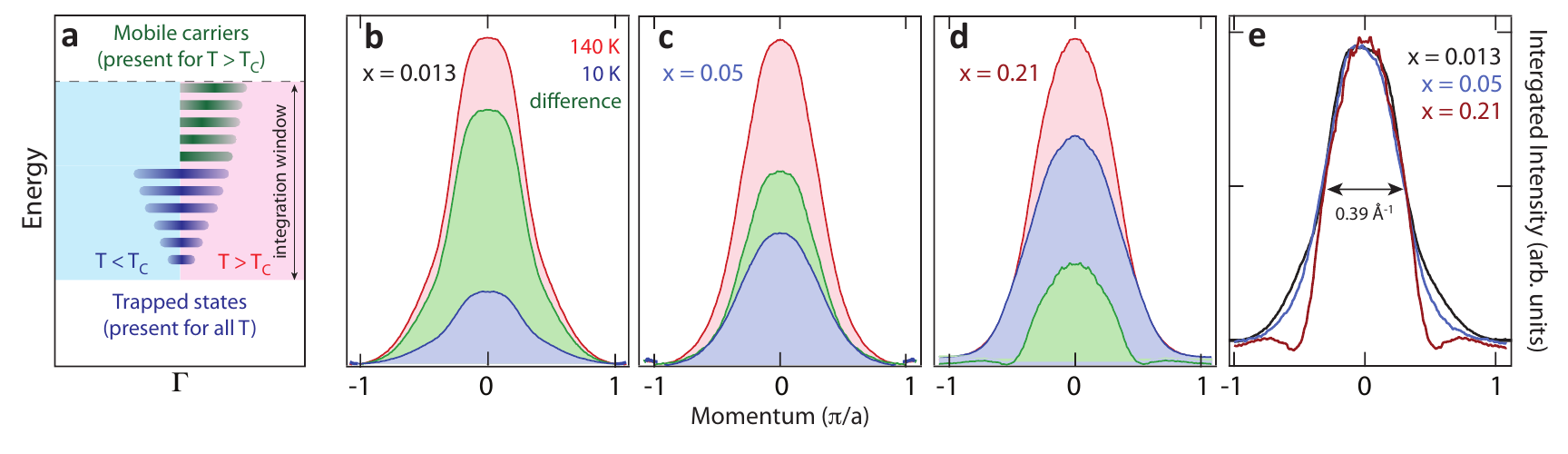}
	\caption{\label{fig:mdc}Momentum distribution curves for doped EuO. (a) Illustration of the electrically active (green) and inactive (blue) states above and below $T_{C}$. (b-d) Integrated MDC curves from -1 eV to +0.1 eV, at 140 K (red) and 10 K (blue), and the difference (green) for (b) $x=0.013$, (c) 0.05, and (d) 0.21.  Each spectrum has been normalized to the peak in the Eu 4$f$ valence band. (e) Summary of difference curves normalized to their peak intensity, illustrating identical widths irrespective of doping value/$k_\mathrm{F}$}  
\end{figure*}

Having established a simple model which accounts qualitatively for the features of our ARPES data, we now proceed with a more quantitative analysis of our experimental data.  Prior ARPES data \cite{shai:2012} has revealed two species of charge carriers coexisting around $\Gamma$, which can be described as 1) electrically active carriers which participate in the temperature dependent metal-insulator transition [green in Fig. \ref{fig:mdc}(a)], and 2) electrically inactive or deeply bound electrons which remain trapped at all temperatures [blue in Fig. \ref{fig:mdc}(a)]. The ratio between these two different states is also approximately consistent with the ratio of the free carriers deduced by Hall measurements versus the percentage of doped Gd cations, on the order of 50\%. ~\cite{mairoser:2010, shai:2012} Nonetheless, an exact equality between the Gd concentration and $k_\mathrm{F}$ cannot be directly established, due to uncertainties in the three-dimensional electronic structure in the $k_{z}$ direction. We note that the model Hamiltonian only considers the electrically active species of charge carrier. Therefore, to isolate the relevant electrically active carriers from the deeply bound states (which are not considered in this work), we subtract the integrated spectral weight at low temperatures around $\Gamma$ from the data measured at 140 K, the results of which are shown in Figs. \ref{fig:mdc}(b-d). We follow this procedure because below $T_{\rm C}$, the mobile carriers are transferred to the $X$ point (the majority spin conduction band minimum), while the deeply bound states are present at all temperatures; therefore, the difference in intensity above and below T$_{\rm C}$ should represent solely the mobile carrier contribution above $T_{\rm C}$.

In Fig. \ref{fig:mdc}(e), we plot the subtracted MDCs (green) for all three doping values on the same axes, each normalized to their maximum intensity. While there are subtle intensity variations at larger $k$, it is rather striking that the full width at half maximum (FWHM) of each curve is essentially identical (0.39 $\pm$ 0.03 \AA$^{-1}$), despite an expected significant overlap of the impurity wave functions and having $k_\mathrm{F}$'s which differ by as much as a factor of 2 [Fig. \ref{fig:SpectralFunctionEvolution}(d)].  This result is strongly suggestive that there is a momentum-dependent suppression of the spectral function, as predicted by our model Hamiltonian in Eq.~\eqref{eq:stot}. The uniform momentum-space width described above corresponds to a value of $r_0 = 4.3$ \AA\ and a real space FWHM of $\Delta r = 10$ \AA, which is remarkably close to the value for $a_B$ = 12 \AA, predicted using simple Thomas-Fermi screening, where $a_Bn_c^{1/3} \geq \frac{1}{4}$ \cite{ashcroft:1976} and $n_c\sim1\times10^{19}$ cm$^{-3}$ for EuO. Given the fcc EuO lattice with 12-fold coordination and an Eu-Eu nearest neighbor distance of 3.6 \AA, the donor extends over multiple lattice sites, consistent with our envelope of intensity in $k$-space. These observations are also consistent with recent atomically resolved scanning tunneling microscopy (STM) measurements on EuO$_{1-x}$ thin films by \textcite{klinkhammer:2014}, who showed an oxygen vacancy was found to have a real space extent with FWHM $\sim$ 7.5 \AA.  Considering that STM measures the square modulus of a real space wave function, $\left|\psi(x)\right|^2$, this would imply a wave function parameterized by $\Delta r \approx 11$ \AA, in excellent agreement with our $k$-space measurements.


While our measurements indicate that the doped carriers form semi-localized states, they do not elucidate the mechanism of this localization. In conventional semiconductors, electrons are localized by the impurity-site potential in a dielectric background. Within the effective mass approximation, this results in a localization length given by the crystal Bohr radius, $a_{B} = \kappa_{0}\hbar^{2}/m^{\ast}e^{2}$. While this description is not formally valid in the present case, a rough estimate can still be obtained. The static dielectric constant of undoped EuO at room temperature is $\kappa_0 = 23.9$,\cite{LandoltBornstein} while the effective mass of EuO has been previously estimated by numerous groups with results ranging from $m^{\ast}=0.35-1.1\ m_e$.~\cite{schoenes:1974,bebenin:1985,patil:1981}
Using these values, we calculate $a_B \approx 10-40$ \AA, comparable to our previous estimates.
The correlated nature of EuO and the localized $f$ moments could also result in a different source of localization. For example, much attention has been devoted to the idea of magnetically induced localization (i.e. magnetic polarons) forming at high temperatures,\cite{torrance:1972,mauger:1983} where the electron is localized owing to its coupling to a disordered magnetic background, although the accuracy of this picture has been questioned recently by \textcite{monteiro:2013}. 

Recently, localized impurity states were also observed for the case of the wide-gap semiconductor Si-doped $\beta$-Ga$_3$O$_2$ in both STM \cite{iwaya:2011} and ARPES \cite{richard:2012}. Similarly, a consistent picture of momentum space and real space was obtained. Nonetheless, the states observed showed no sign of dispersion in ARPES. Moreover, the localized states appeared below the expected band minimum, consistent with the effective mass description, while in EuO, the localized states appear around the $\Gamma$ point, with the (conduction) band minimum located at the Brillouin-zone boundary.

In conclusion, we have investigated the semiconductor-to-metal transition in Eu$_{1-x}$Gd$_x$O through the evolution of the electronic spectral weight obtained by ARPES, finding a dispersing pocket around the $\Gamma$ point which grows with increasing electron doping $x$. Comparing to an effective model describing dilute, quasi-localized impurity states allows us to not only gain momentum-resolved spectral information from ARPES measurements on doped EuO, but also to extract the length scale related to the semiconductor-to-metal transition in this material, demonstrating the validity of this simple model even in strongly correlated materials. This length scale of $\Delta r = 10$ \AA{} does not change with increasing doping, providing a clear signature of the Bohr radius of the previously localized impurity states, even well into the metallic regime. 

This work was supported by the National Science Foundation through DMR-0847385, DMR-1308089, and through the Materials Research Science and Engineering Centers program (DMR-1120296, the Cornell Center for Materials Research), and the Research Corporation for Science Advancement (2002S). This work was performed in part at the Cornell NanoScale Facility, a member of the National Nanotechnology Infrastructure Network, which is supported by the National Science Foundation (Grant No. ECCS-0335765). M.H.F acknowledges support from the Swiss Society of Friends of the Weizmann Institute of Science.

\bibliography{refs}

\end{document}